\documentclass{webofc}
\usepackage[varg]{txfonts}   
\usepackage{hyperref}
\usepackage{url}
\hypersetup{colorlinks=true,citecolor=blue,urlcolor=blue,linkcolor=blue}
\usepackage{subcaption}

\begin{document}
\title{Application of sequence learning for predicting radiation damage of the CMS electromagnetic calorimeter}

\author{
\firstname{Mario Ivan} \lastname{Gallegos Torres}\inst{1}
\and \firstname{Leonid} \lastname{Serkin}\inst{1}\fnsep\thanks{\email{lserkin@ciencias.unam.mx}}
\and \firstname{Guy} \lastname{Paic}\inst{2}
}

\institute
{Facultad de Ciencias, Universidad Nacional Autónoma de México, Circuito Exterior s/n, Ciudad de México 04510, México. 
\and
Instituto de Ciencias Nucleares, Universidad Nacional Autónoma de México, Apartado Postal 70-543, Ciudad de México 04510, México.
}

\abstract{In this paper we use machine learning methods to predict radiation damage in the lead tungstate crystals of the CMS electromagnetic calorimeter at the Large Hadron Collider. We analyze LHC Open Data collected from 2016 to 2018 and study the time evolution of the crystal optical transparency. We apply deep neural network models to predict its behavior over different future time intervals, and find that encoder-decoder sequence-to-sequence architectures can effectively describe crystal aging.
}
\maketitle

\section{Introduction}  

The Compact Muon Solenoid (CMS) experiment at CERN's Large Hadron Collider (LHC) is designed to detect and measure particles produced in proton collisions~\cite{CMS}. One of the key components of the CMS detector is the electromagnetic calorimeter (ECAL), which is composed of lead tungstate (PbWO$_4$) crystals and is used to measure the energy of electrons and photons generated in collisions at the LHC~\cite{TDR}. This detector played a key role in the discovery of the Higgs boson in the two-photon decay channel in 2012~\cite{Higgs1}, as well as in the precise measurement of the Higgs boson mass and in many other measurements~\cite{CMSPub}.

\begin{figure}
    \centering
    \begin{subfigure}[t]{0.48\textwidth}
        \centering
        \includegraphics[width=\textwidth,height=5cm,keepaspectratio]{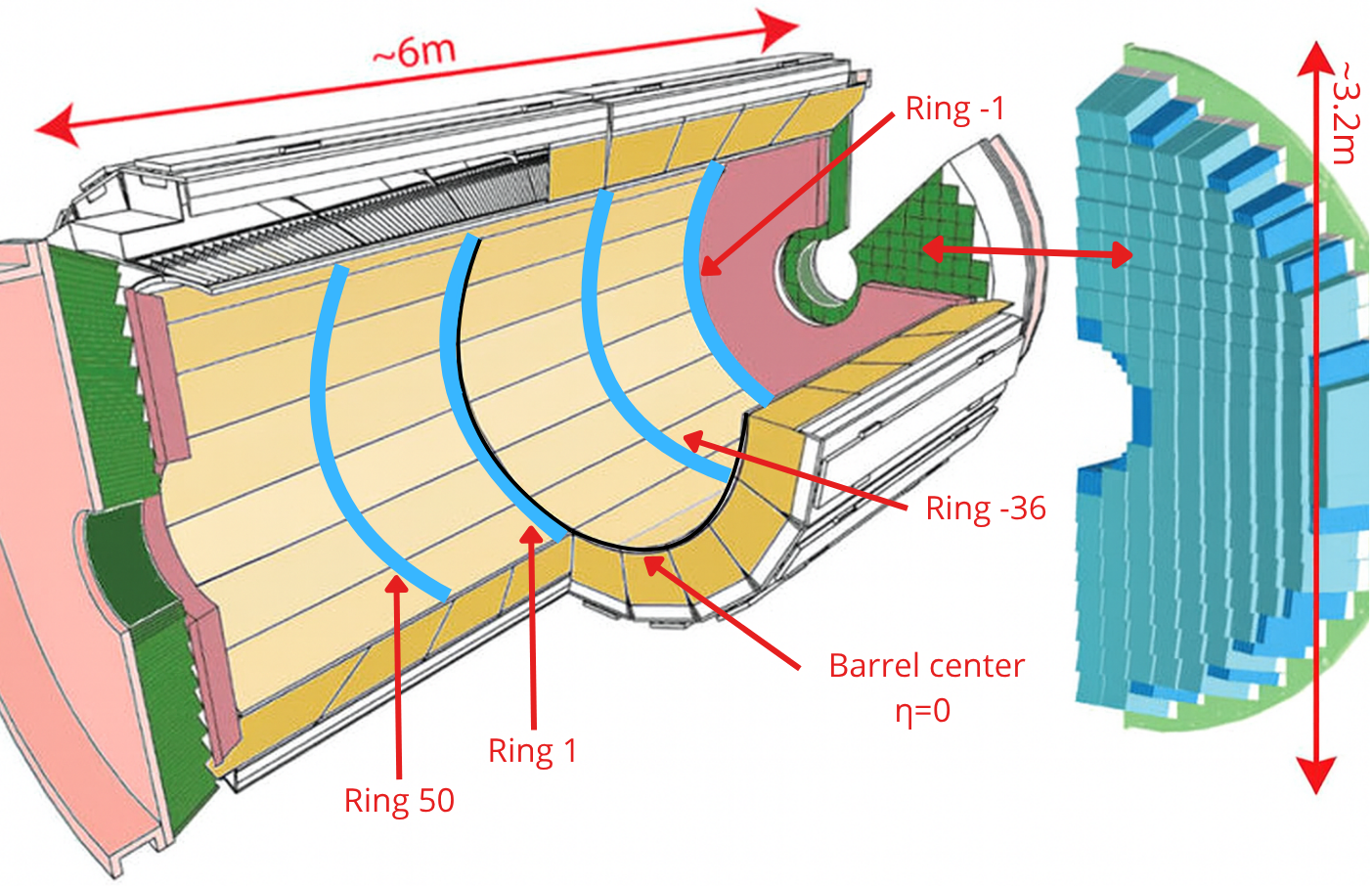}
        \caption{}
        \label{fig:fig1a}
    \end{subfigure}
    \hfill
    \begin{subfigure}[t]{0.48\textwidth}
        \centering
        \includegraphics[width=\textwidth,height=5cm,keepaspectratio]{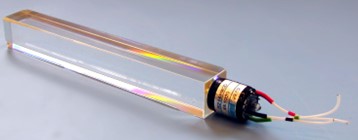}
        \caption{}
        \label{fig:fig1b}
    \end{subfigure}
    \caption{(a) Schematic of the ECAL system of the CMS experiment at the LHC. (b) PbWO$_4$ crystals used in the CMS ECAL.}
    \label{fig:FIG1}
\end{figure}

\medskip
\noindent
The ECAL, shown schematically in Fig. \ref{fig:fig1a}, consists of 75,848 lead tungstate crystals, each approximately 22~cm long, with a cross-sectional area of 2 $\times$ 2~cm$^2$ and a weight of about 1~kg. These crystals, shown in Fig. \ref{fig:fig1b}, are optically transparent and emit short pulses of light when they absorb ionizing radiation. When a high-energy particle (an electron or photon) enters a crystal, it collides with heavy nuclei, producing a cascade of secondary electrons and photons. These electrons excite nearby atoms; as these return to their original energy level, they emit blue light, which is detected by a photomultiplier. The amount of blue light emitted is proportional to the amount of energy deposited in the crystal by the initial particle.

\medskip
\noindent
Due to the intense radiation during LHC operation, the optical transparency of the crystals decreases progressively over time. This degradation phenomenon, known as ``radiation damage,'' is associated with the creation of atomic-level defects in the crystals, which act as color centers by absorbing the light that would otherwise propagate through the material~\cite{CMSCalib}.

\medskip
\noindent
The accuracy of electron and photon detection in the ECAL depends on the ability to measure and correct changes in crystal transparency. For this purpose, the CMS experiment uses a laser monitoring system that measures, every 40 minutes, changes in the optical properties of the crystals caused by the radiation generated in collisions. Since predicting the behavior of optical transparency is essential for the integrity of the experimental measurements and the physical results derived from them, it is crucial to anticipate and correct the variations caused by radiation damage~\cite{CMSCalib}.

\section{Deep neural networks for predicting radiation damage}  
\medskip
\noindent
Deep Neural Networks (DNNs) are extremely powerful machine learning (ML) models that achieve excellent performance on difficult problems such as speech recognition and visual object recognition, and have been widely used in the High Energy Physics community~\cite{chollet}. Motivated by recent work by Joshi \textit{et al.}~\cite{jos}, we analyze the temporal evolution of crystal optical transparency and explore the application of DNN-based models to predict radiation damage over different future time intervals.

\begin{figure}[h]
    \centering
    \includegraphics[width=0.6\linewidth]{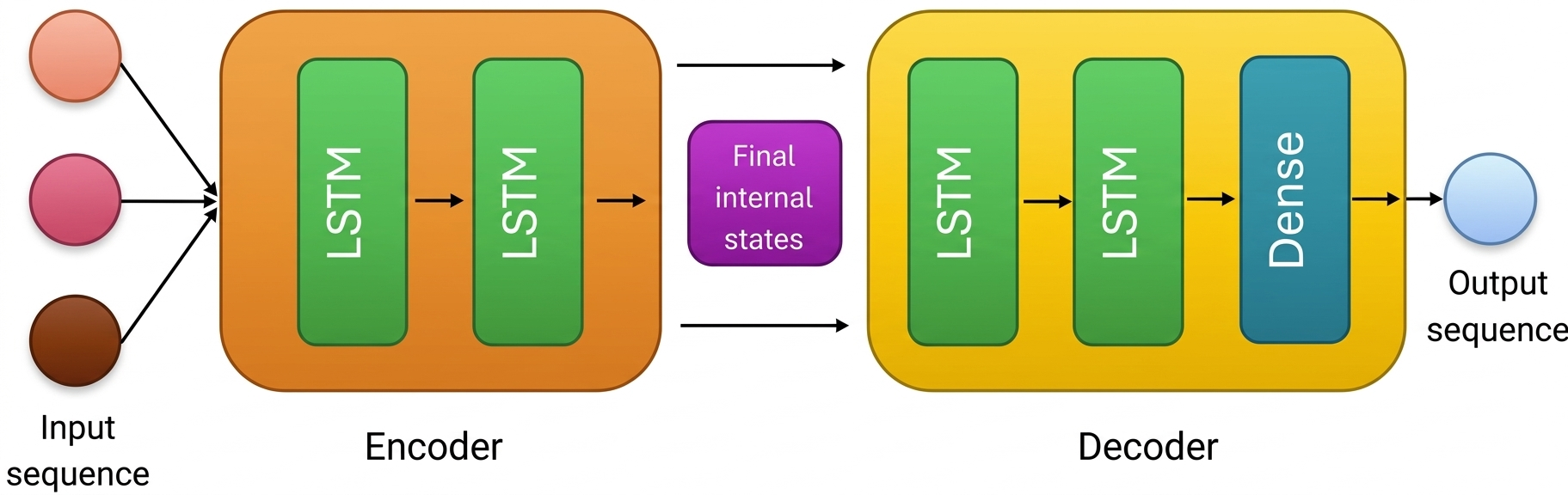}
    \caption{Seq2Seq architecture used in this work.}
    \label{fig:SEQ}
\end{figure}

\medskip
\noindent
The main focus of this paper is on encoder--decoder sequence-to-sequence (Seq2Seq) architectures~\cite{Sutskever}. The Seq2Seq architecture used in this work consists of an encoder--decoder structure based on stacked Long Short-Term Memory (LSTM) layers, as shown in Fig.~\ref{fig:SEQ}. The encoder processes the input sequence and compresses the temporal information into its final hidden and cell states, capturing long-term dependencies. These states are used to initialize the decoder, which generates the output sequence. Starting from the last observed value, the decoder produces predictions iteratively through a dense layer, feeding each predicted value back as input for the next time step until the desired prediction horizon is reached.

\medskip
\noindent
The Seq2Seq model employs two  LSM layers with 512 units each in both the encoder and decoder, and is trained for 400 epochs using teacher forcing, with ``MinMaxScaler'' normalization. The minimization is performed using the Adam algorithm, and the software implementation is carried out in TensorFlow.

\medskip
\noindent
The publicly available LHC Open Data dataset hosted on the Zenodo platform~\cite{zen} was used. This dataset contains the laser response of PbWO$_4$ crystals collected during Run 2 of the LHC (2016--2018). Each record includes detailed information on the crystal identification, precise timestamps of the last calibration (transparency) value, as well as accurate data on the integrated luminosity delivered since the previous measurement.

\medskip
\noindent
After cleaning and exploring the data, we select several representative ECAL crystals for detailed study. The problem is formulated as a sequence forecasting task: given a sliding window of past calibration values, we use a Seq2Seq model to predict future calibration values for a given crystal. We train the model on the full 2016 dataset and evaluate it on 2017--2018 data, with the inputs normalized to unit variance.

\medskip
\noindent
The prediction accuracy is evaluated using the mean absolute error (MAE):
\begin{equation}
\mathrm{MAE} = \frac{1}{n}\sum_{i=1}^n \left| y_i^{\rm true} - y_i^{\rm pred} \right|;
\end{equation}
\noindent
and the mean absolute percentage error (MAPE):
\begin{equation}
\mathrm{MAPE} = \frac{100}{n}\sum_{i=1}^n \left|\frac{y_i^{\rm true} - y_i^{\rm pred}}{y_i^{\rm true}}\right|.
\end{equation}

\section{Seq2Seq prediction results}  

\begin{figure}[h]
    \centering
    \begin{subfigure}[t]{0.48\textwidth}
        \centering
        \includegraphics[width=\textwidth,height=5cm,keepaspectratio]{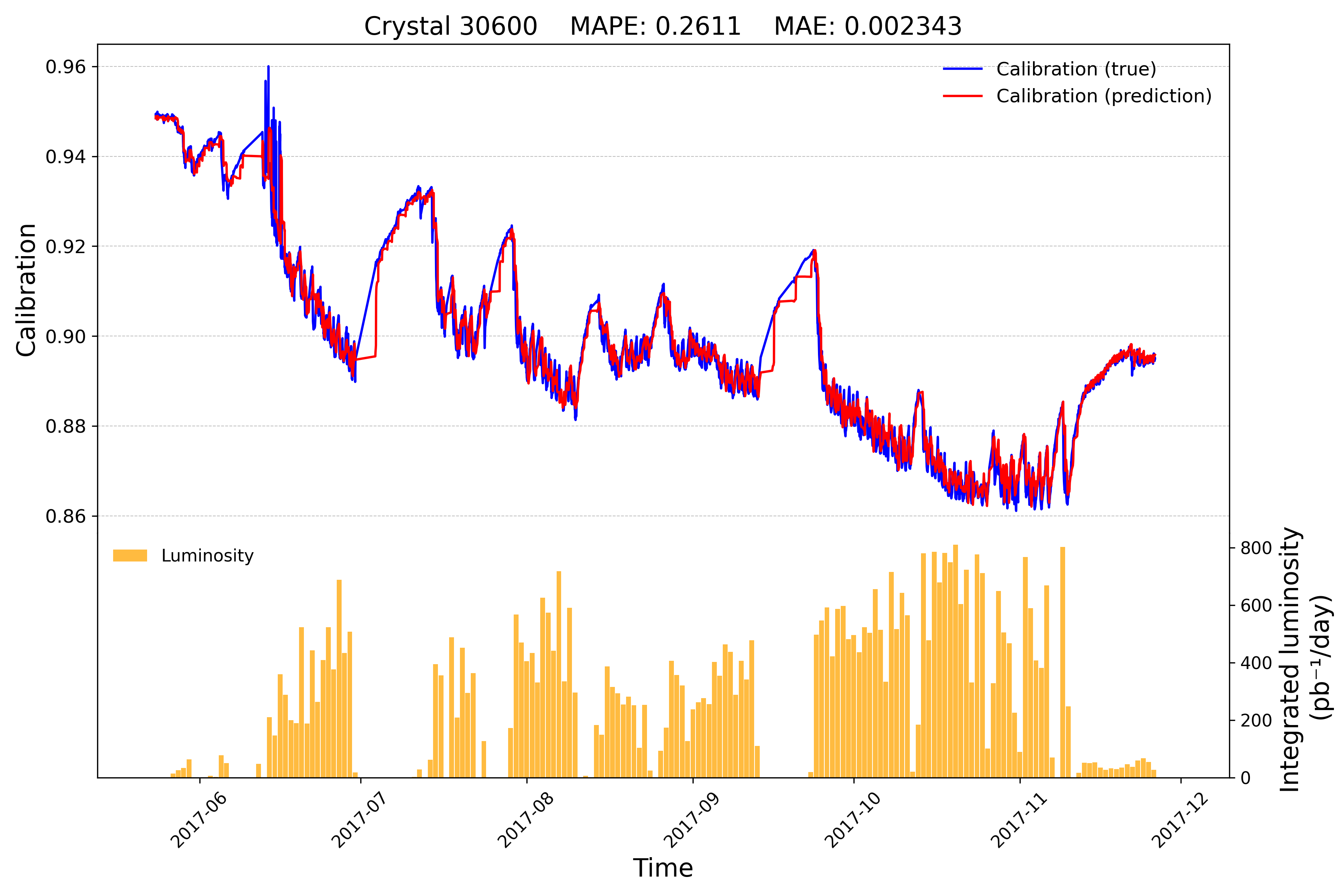}
        \caption{}
        \label{fig:fig2a}
    \end{subfigure}
    \begin{subfigure}[t]{0.48\textwidth}
        \centering
        \includegraphics[width=\textwidth,height=5cm,keepaspectratio]{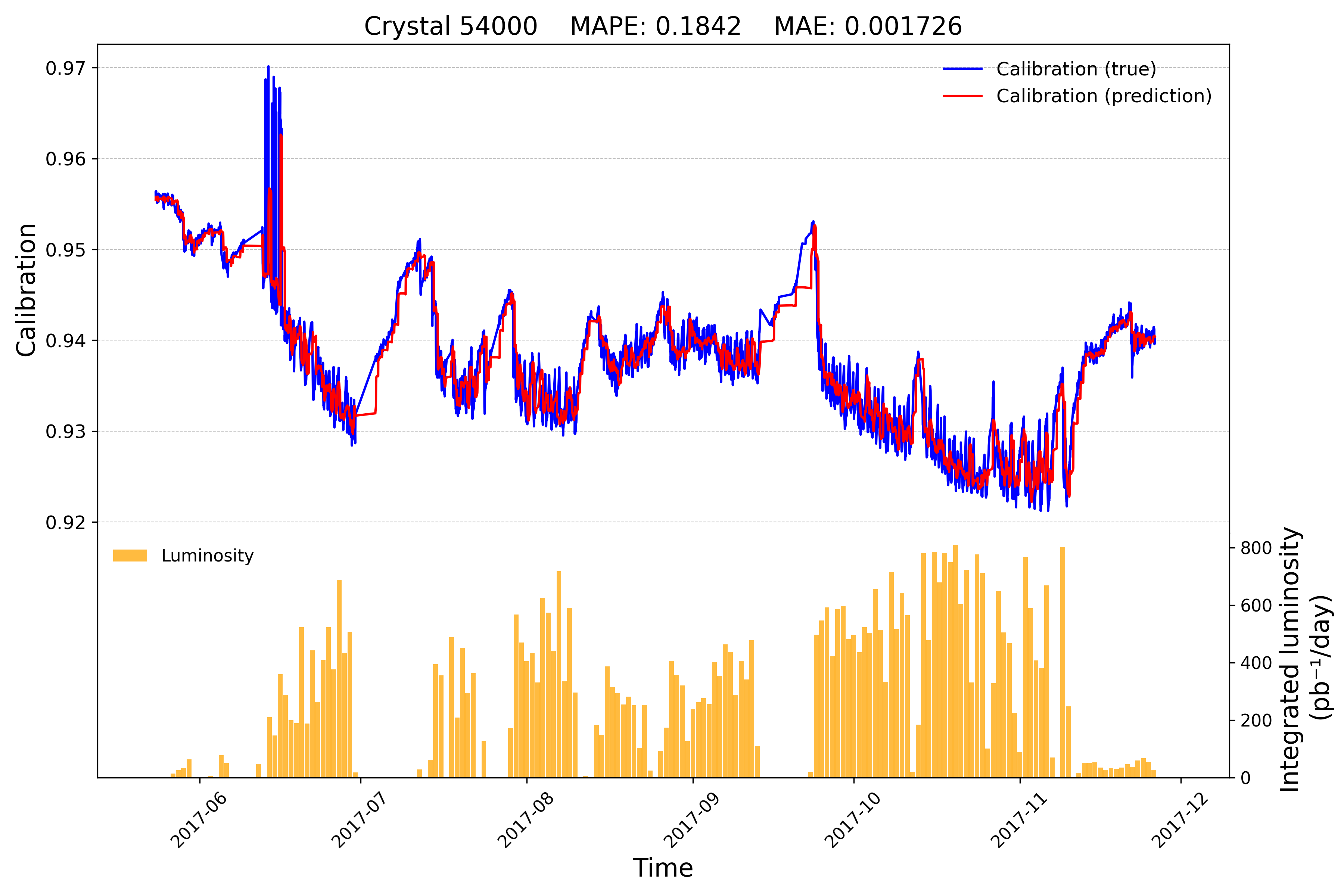}
        \caption{}
        \label{fig:fig2b}
    \end{subfigure}
    \caption{Comparison between the true (blue line) and predicted (red line) calibration for ECAL crystal (a) 30600 and (b) 54000 as a function of time, using a 12-step prediction horizon. The lower panel displays the integrated daily luminosity.}
    \label{fig:FIG2}
\end{figure}

Fig.~\ref{fig:FIG2} presents a comparison between the true and predicted calibration for two ECAL crystals, 30600 and 54000, as a function of time during the 2017 LHC data-taking period, using a 12-step prediction horizon. For reference, the lower panel shows the integrated luminosity. It can be observed that during periods of collisions, the calibration values decrease due to radiation damage, while during periods without collisions, the crystals recover their transparency.

\medskip
\noindent
The model achieves a MAE of $2.3 \times 10^{-3}$ and a MAPE of $\approx 26$\% for crystal 30600, and an MAE of $1.7 \times 10^{-3}$ with a MAPE of $\approx 18$\% for crystal 54000, showing a good agreement with the true calibration values in both cases.

\begin{figure}[h]
    \centering
    \begin{subfigure}[t]{0.48\textwidth}
        \centering
        \includegraphics[width=\textwidth,height=5cm,keepaspectratio]{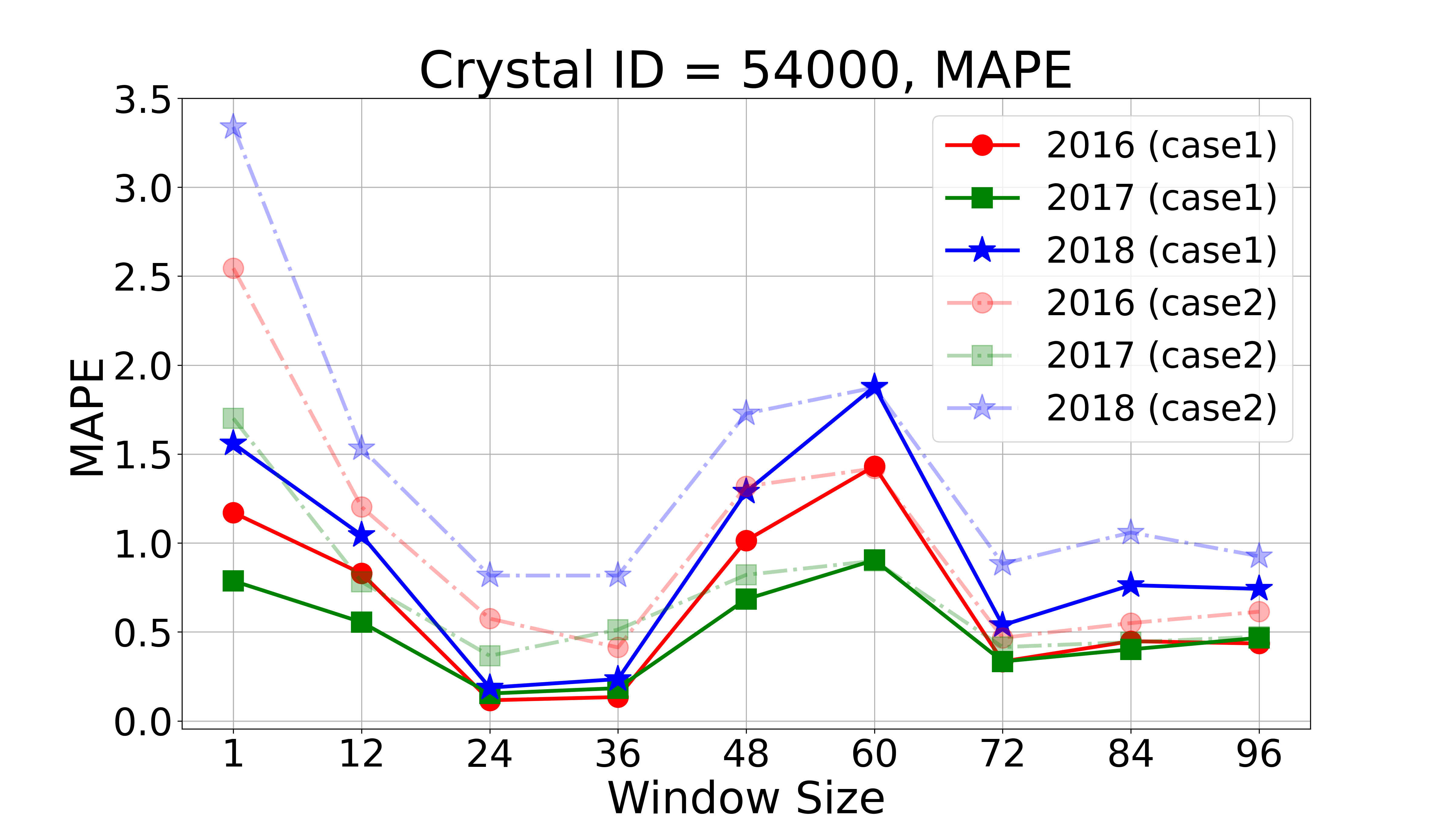}
        \caption{}
        \label{fig:fig9a}
    \end{subfigure}
    \hfill
    \begin{subfigure}[t]{0.48\textwidth}
        \centering
        \includegraphics[width=\textwidth,height=5cm,keepaspectratio]{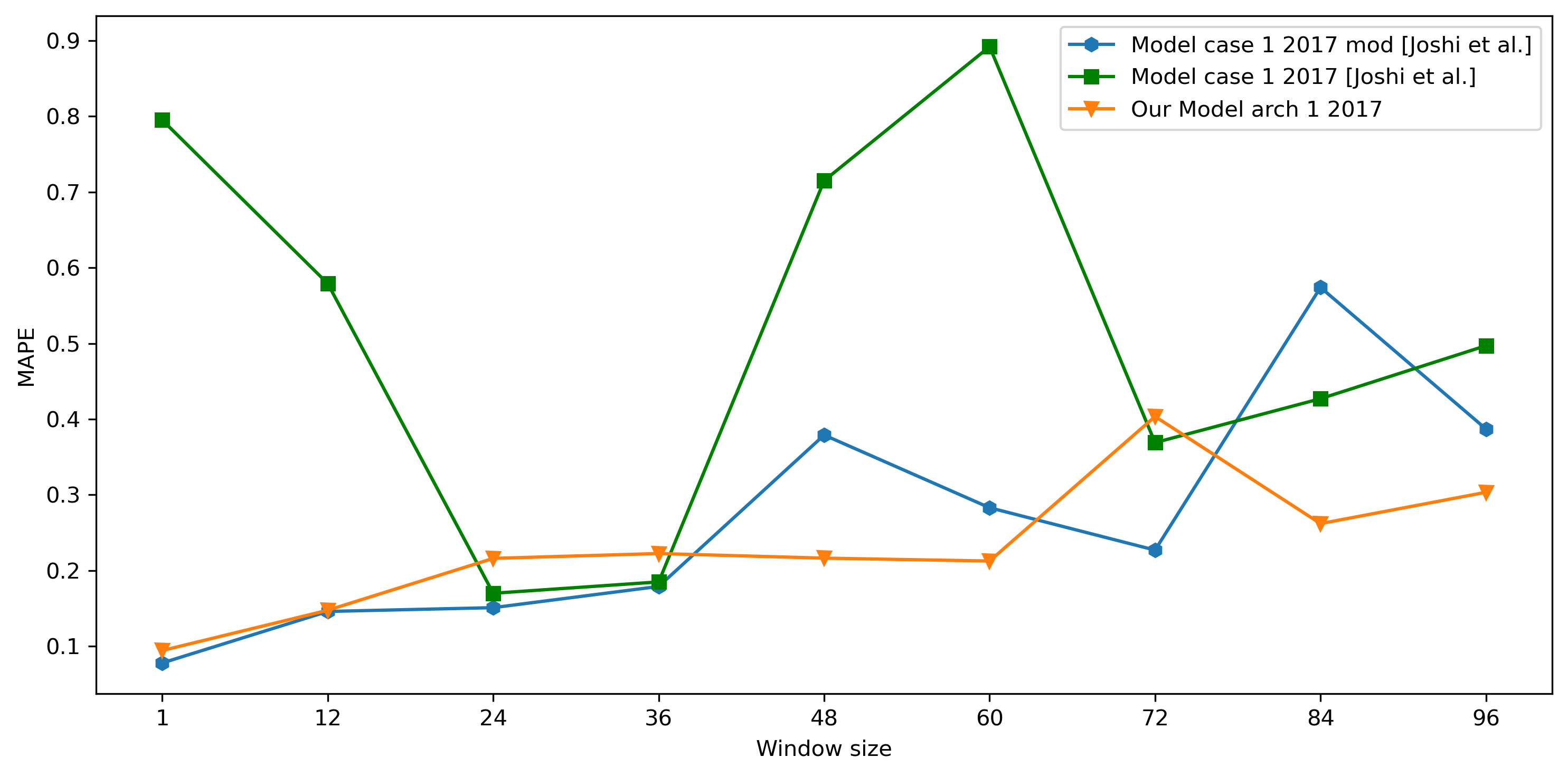}
        \caption{}
        \label{fig:fig9b}
    \end{subfigure}
    \caption{
    (a) Results reported by Joshi \textit{et al.}~\cite{jos}, including the original 2017 Case 1 (green). (b) Comparison of this result (green) with the corrected code implementation (blue), and comparison with the Seq2Seq model (yellow) developed in this paper. All results are shown for ECAL crystal 54000.
    }
    \label{fig:fig9}
\end{figure}

\medskip
\noindent
In the following, a comparison is made for predictions on ECAL crystal 54000 using the model from Ref.~\cite{jos, crystal}. Fig.~\ref{fig:fig9a}, taken from Ref.~\cite{jos}, shows a surprising dip in MAPE at intermediate window sizes. We traced this behavior to a coding bug in the original author's implementation. Firstly, we reproduced their model shown as green curve in Fig.~\ref{fig:fig9b}. Afterwards, we obtained the corrected results, shown as blue line in Fig.~\ref{fig:fig9b}, noticing that the difference between these curves highlights the impact of the bug on the reported performance in Ref.~\cite{jos}.

\medskip
\noindent
Finally, the yellow curve in Fig.~\ref{fig:fig9b} corresponds to the predictions obtained with our Seq2Seq model. Our independent implementation (with identical architecture) exhibits a different and more stable behavior across window sizes, with MAPE varying smoothly with input length and showing no anomalous dip. The best performance is achieved at short horizons (MAPE $\sim$9--15\%), while intermediate window sizes maintain a stable error of $\sim$20\%. Although performance degrades for very large windows, the model remains significantly more stable and predictable than previous approaches, indicating improved generalization.

\clearpage

\section{Conclusions}  
In this work, we presented a Seq2Seq LSTM model for forecasting the transparency of CMS ECAL crystals. The model, based on an encoder--decoder architecture trained with teacher forcing, is able to capture the temporal evolution of radiation damage and recovery effects observed during LHC operation.

\medskip
\noindent
Our results show that the Seq2Seq architectures can model the long-term time evolution of crystal aging and our Seq2Seq model provides good agreement with the measured calibration values. In particular, the model performs best at short prediction horizons, reaching MAPE values of $\sim$9--15\%, while maintaining stable performance of $\sim$20\% for intermediate window sizes. In addition, we compare our results with previous work and find a coding bug in the original implementation, which leads to anomalous behavior in the reported results.

\medskip
\noindent
Future work will focus on extending this approach to a larger set of crystals, including full ECAL rings, and on exploring alternative deep learning models such as multi-output architectures. Since the response may vary from one crystal to another, a systematic study will be needed to determine how well models trained on individual crystals generalize to crystals with different aging and recovery patterns and to complete ECAL rings, as well as to explore more advanced sequence-learning strategies. By relying on publicly available data and reproducible methodologies, this work provides a solid basis for further developments in data-driven calibration studies. These predictions can be useful for the CMS Collaboration in improving calibration strategies and reducing the impact of radiation damage on detector performance.


\begin{thebibliography}{}


\bibitem{CMS}
CMS Collaboration, \textit{Performance of the CMS electromagnetic calorimeter in $pp$ collisions at 13 TeV}. JINST \textbf{19} (2024) P09004. \url{https://doi.org/10.1088/1748-0221/19/09/P09004}

\bibitem{TDR}
CMS Collaboration, \textit{CMS physics: Technical design report volume 1: Detector performance and software}, CERN-LHCC-2006-001, 2006. \url{https://inspirehep.net/literature/1614070}

\bibitem{Higgs1}
CMS Collaboration, \textit{Observation of a new boson at a mass of 125 GeV with the CMS experiment at the LHC}. Phys. Lett. \textbf{B716} (2012) 30. \url{https://doi.org/10.1016/j.physletb.2012.08.021}

\bibitem{CMSPub}
CMS Collaboration. Scientific Results. 2026. \url{https://cms.cern/org/cms-scientific-results}.

\bibitem{CMSCalib}
CMS Collaboration, \textit{Energy calibration and resolution of the CMS electromagnetic calorimeter in $pp$ collisions at 7 TeV}, JINST \textbf{8} (2013) P09009. \url{https://doi.org/10.1088/1748-0221/8/09/P09009}

\bibitem{chollet}
K. Albertsson et al., \textit{Machine Learning in High Energy Physics Community White Paper}, J. Phys. Conf. Ser. 1085 (2018) 2, 022008. \url{https://doi.org/10.48550/arXiv.1807.02876}

\bibitem{jos}
B. Joshi, T. Li, B. Liang, R. Rusack, \& J. Sun, \textit{Predicting the future of the CMS detector: Crystal radiation damage and machine learning at the LHC}. arXiv:2303.15291 [physics.ins-det] (2023). \url{https://arxiv.org/abs/2303.15291}


\bibitem{Sutskever}
I. Sutskever, O. Vinyals and Q. V. Le, \textit{Sequence to Sequence Learning with Neural Networks}, arXiv:1409.3215 [cs.CL] (2014). \url{https://arxiv.org/abs/1409.3215} 


\bibitem{zen}
B. Joshi and R. Rusack, \textit{Laser Response in ECAL Crystals in CMS Detector}. Zenodo, 2022. \url{https://doi.org/10.5281/zenodo.7510572}

\bibitem{crystal}
FAIR-UMN, fair\_ecal\_monitoring. \url{https://github.com/FAIR-UMN/fair_ecal_monitoring}



\end{thebibliography}
\end{document}